\documentclass[11pt]{revtex4}
\usepackage{graphicx}
\begin{document}
\title{Entanglement Entropy at Finite Density from Extremal Black Holes}
\author{Brian Swingle}
\affiliation{Department of Physics, Massachusetts Institute of Technology, Cambridge, MA 02139}
\email{bswingle@mit.edu}
\begin{abstract}
I compute the entanglement entropy of a strongly coupled $2+1$d quantum field theory containing fermions at finite density using gauge/gravity duality.  The dual geometry is an extremal black hole in $3+1$d Einstein-Maxwell theory.  This system was recently shown to exhibit non-Fermi liquid behavior, but the leading geometrical contribution to the entanglement entropy does not produce an expected violation of the boundary law.  I discuss this negative result in the context of attempts to find highly entangled states of quantum matter.
\end{abstract}
\maketitle
\section{Introduction}
Entanglement is a still mysterious substance or property of quantum systems that distinguishes them from their classical counterparts.  In quantum many body physics, by analogy with the long range correlations present in symmetry breaking phases, some quantum phases possess what has been termed long range entanglement.  Though theorists know how to give meaning to this phrase in certain cases, the general principles are still murky.  The class of long range entangled phases includes fractional quantum hall systems, spin liquids, and many others.  It is therefore interesting to develop methods for describing the entanglement structure of many body systems.

One widely used tool for this purpose is entanglement entropy.  To compute the entanglement entropy we divide a many body system $\mathcal{S}$, assumed to be in a pure state, into two sub-systems, $\mathcal{R}$ and $\mathcal{S}\backslash \mathcal{R}$.  The von Neumann entropy $S_{\mathcal{R}}$ of the reduced density matrix $\rho_{\mathcal{R}}$ is the entanglement entropy.  The utility of $S_{\mathcal{R}}$ is that it vanishes when $\rho_{\mathcal{R}}$ is pure, in other words, when $\mathcal{R}$ and $\mathcal{S}\backslash \mathcal{R}$ are unentangled.  We interpret $S_{\mathcal{R}}$ as measuring the amount of entanglement between $\mathcal{R}$ and $\mathcal{S}\backslash \mathcal{R}$.

The basic behavior of $S_{\mathcal{R}}$ in a local many body system is captured by the boundary law \cite{arealaw1}.  In many situations the entanglement entropy $S_{\mathcal{R}}$ is found to scale as the boundary $\partial \mathcal{R}$ of region $\mathcal{R}$.  For example, a gapped bosonic system in three spatial dimensions with $\mathcal{R}$ a ball of radius $L_{\mathcal{R}}$ has an entanglement entropy that scales as $L_{\mathcal{R}}^2$.  An important exception to this rule is free fermions at finite density where the entanglement entropy scales as $S_{\mathcal{R}} \sim L_{\mathcal{R}}^{d-1} \ln L_{\mathcal{R}}$.  It would be interesting to understand other examples of highly entangled states that violate the boundary law.

Strongly interacting many body systems are natural candidates for exotic entangled states. The AdS/CFT correspondence, or more generally gauge/gravity duality \cite{magoo}, is a relatively new and powerful method for investigating properties of strongly interacting systems.  Within this framework, quantum field theories without gravity are dual to higher dimensional theories of quantum gravity in asymptotically anti-de Sitter space.  The correspondence has attracted attention in condensed matter physics because it answers interesting questions about quantum many body systems that are otherwise difficult to address.  For example, computing the entanglement entropy is complicated within the field theory, but on the gravity side it reduces to a minimal surface calculation in the classical limit \cite{holo_ee,holo_ee_f}.  Holographic calculations of the entropy have been carried out in a number of cases \cite{holo1,holo2,holo3,holo4}.

Here I compute the entanglement entropy via gauge/gravity duality for a $2+1$d quantum field theory.  The field theory at finite density and zero temperature is dual to a $3+1$d extremal charged black hole.  This system is interesting because of recent work demonstrating the existence of a non-Fermi liquid phase of fermions with a sharp Fermi surface but no sharp quasiparticle \cite{nfl1}.  The system also displays an emergent quantum critical behavior that arises from the near horizon AdS$_2$ geometry of the extremal black hole \cite{nfl2}.  Some heuristic considerations suggest that the entanglement entropy may have an extra logarithmic divergence due to the near horizon geometry.  This suspicion is supported by a recent proposal that non-Fermi liquids should violate the boundary law \cite{fee}.  However, the conclusion of this work is that no such violation exists at the geometrical level.  The strongly coupled non-Fermi liquid phase may violate the boundary law, but this violation is not present in the bare geometry.  After this work was completed, I learned of previous work \cite{scooped} with different motivations in which some of my results were obtained.

\section{Black hole geometry}
I consider Einstein-Maxwell theory in $3+1$d.  The action for an abelian gauge field $A_M$ coupled to Einstein gravity $g_{MN}$ is given by
\begin{equation}
S = \frac{1}{2 \kappa^2} \int d^4\, x \sqrt{-g} \left[ R + \frac{6}{L^2} - \frac{L^2}{g_F^2} F^2 \right]
\end{equation}
where $F_{MN}$ is the electromagnetic field strength and $g_F$ is a dimensionless gauge coupling.  The field equations have a solution corresponding to a charged black hole with mass $M$ and charge $Q$.  The metric is given by
\begin{equation}
ds^2 = \frac{r^2}{L^2} \left( - f dt^2 + dx^2 + dy^2 \right) + \frac{L^2}{r^2 f} dr^2,
\end{equation}
where
\begin{equation}
f(r) = 1 + \frac{Q^2}{r^4} - \frac{M^2}{r^3}, \,\, A_0 = \mu \left(1 - \frac{r_0}{r} \right), \,\, \mu = \frac{g_F Q}{L^2 r_0}.
\end{equation}
The location $r_0$ of the black hole horizon is determined by the equation $f(r_0) = 0$.  The parameter $\mu$ corresponds to a chemical potential for the boundary theory.  Note that we do not know the details of the boundary theory.  We do know that turning on the charged black hole corresponds to putting a finite density of some conserved charge in the boundary theory.

It is convenient to introduce rescaled versions of the parameters appearing in the classical background.  These rescaled quantities (denoted by a tilde) are
\begin{equation}
r = r_0 \tilde{r},\,\, t = \frac{L^2}{r_0} \tilde{t}, \,\, A_0 = \frac{r_0}{L^2} \tilde{A}_0 , \,\, M = r_0^3 \tilde{M}, \,\, Q = r_0^2 \tilde{Q}.
\end{equation}
In terms of rescaled variables the metric becomes
\begin{equation}
ds^2 = L^2 \left( - f d\tilde{t}^2 + d\tilde{x}^2 + d\tilde{y}^2 \right) + \frac{L^2}{\tilde{r}^2 f} d\tilde{r}^2
\end{equation}
with $f(\tilde{r}) = 1 + \tilde{Q}^2 \tilde{r}^{-4} - (1 +\tilde{Q}^2) \tilde{r}^{-3}$.  The dimensionless temperature is given by
\begin{equation}
T = \frac{3 - \tilde{Q}^2}{4 \pi},
\end{equation}
where $\tilde{Q}^2  = 3$ is the zero temperature extremal limit.  I refer only to scaled variables from here on, and I will remove the tildes to simplify notation.

\section{Entanglement Entropy}
To compute the entanglement entropy of the field theory via the AdS/CFT correspondence one must compute the area of a certain minimal surface in the bulk.  The region $\mathcal{R}$ is regarded as living at the boundary of AdS in the ultraviolet.  The minimal surface is then defined by extending the boundary $\partial \mathcal{R}$ living on the conformal boundary of AdS into the bulk.  The entanglement entropy is the area of this surface divided by $4 G_N$ where $G_N$ is the bulk Newton's constant.  This prescription is motivated by the relation between entropy and area for black holes \cite{holo_ee}, and it was proved in the classical super-gravity limit \cite{holo_ee_f}.

Let's first consider the case of pure AdS with no black hole $M = Q = 0$.  I work with an infinite strip geometry where $\mathcal{R}$ is given by $x \in [-L_x/2,L_x/2] \,\, y \in [-L_y/2,L_y/2]$ with $L_y \gg L_x$.  The translation invariance in $y$ simplifies the structure of the minimal surface making the radial coordinate $r$ a function of $x$ alone.  In fact, it is more convenient to use the coordinate $z = 1/r$ to perform the minimal surface computation.  The boundary of AdS is at $z=0$, and all minimal surfaces are required to match onto the curve $\partial \mathcal{R}$ at $z=0$.  In terms of the $z$ coordinate the metric of AdS$_4$ is
\begin{equation}
ds^2 = \frac{1}{z^2} \left( dz^2 + dx^2 + dy^2 - dt^2 \right),
\end{equation}
and the area of a surface $z(x,y) = z(x)$ is
\begin{equation}
A = \int dx dy \frac{1}{z^2}\sqrt{1 + \dot{z}^2},
\end{equation}
where $\dot{z} = dz/dx$.

The goal is to minimize $A$ subject to the boundary condition that the surface terminates on the one dimensional boundary of $\mathcal{R}$ at $z = 0$ corresponding to the conformal boundary of AdS.  Since $A$ does not depend explicitly on $x$ there is a conserved quantity given by
\begin{equation}
\frac{1}{z^2} \frac{1}{\sqrt{1 + \dot{z}^2}} = \frac{1}{a^2}.
\end{equation}
By symmetry the minimal surface will be a symmetric function of $x$ so that $\dot{z} = 0$ occurs at $x=0$ where the minimal surface achieves its maximum depth $z = a$.  The relationship between $a$ and the boundary condition $L_x$ is given by
\begin{equation}
L_x = 2\int_0^a dz \frac{z^2}{\sqrt{a^4 - z^4}}.
\end{equation}
Rescaling the this integral gives $a$ proportional to $L_x$.

Using the conserved quantity the area of the minimal surface can be written in terms of $z$ alone as
\begin{equation}
A = 2 L_y \int_0^a dz \frac{a^2}{z^2 \sqrt{a^4 - z^4}}.
\end{equation}
This integral is divergent at $z=0$ corresponding to an ultraviolet singularity in the dual field theory.  This cutoff dependent term is nothing but the leading non-universal boundary law term common to local quantum systems.  The AdS/CFT correspondence thus predicts that conformal field theories have a leading non-universal boundary law term in the entanglement entropy in more than one spatial dimension.  A similar divergence occurs in the black hole background, but I am only interested in infrared contributions to the entropy.

\section{Black Hole Entropy}
Now let's consider the block hole geometry with finite charge and mass.  The space-time no longer has scale symmetry, but for $a \ll 1$ the pure AdS results remain approximately valid.  As the minimal surface begins to explore the black hole horizon, interesting new effects appear.  It is known that for generic black holes the minimal surface can sit on the horizon providing an extensive (from the boundary point of view) contribution to the entropy.  This part of the minimal surface represents the thermal entropy rather than the entanglement entropy of the region $\mathcal{R}$.  However, the extremal black hole is a special case.

Finite temperature black holes have a first order zero in the emblackening factor $f$ at the horizon.  In the extremal limit the black hole develops a second order zero in $f$ at the horizon.  This difference is important because the horizon tends to repel the minimal surface.  What does this higher order zero signify physically?  In the Poincare patch of pure AdS, the ``renormalization group distance" $\int^{z_0}_\epsilon \frac{dz}{z}$, diverges as $z_0$ approaches the Poincare horizon.  The divergence signifies the presence of excitations down to zero energy.  On the other hand, the distance $\int^1_\epsilon \frac{dz}{z \sqrt{f}}$ to a finite temperature black hole horizon at $z=1$ is finite.  This indicates roughly that low energy modes have disappeared at finite temperature, although there may still be a few quasinormal collective modes.  However, the story changes in the extremal case, and the renormalization group distance now diverges as the horizon is approached.  Again, the divergence indicates the presence of excitations at low energy.  These modes are associated with the near horizon AdS$_2$ region and emergent quantum criticality \cite{nfl2}.  In fact, the story is similar to the way the AdS/CFT correspondence originally appeared from the near horizon limit of extremal black branes in $10$d supergravity.

By analogy with the finite temperature black hole, the simplest possibility for the minimal surface consists of a surface, call it $\Sigma_1$, that drops vertically from the boundary to the horizon and then runs along it.  However, this surface cannot sit exactly on the horizon because the double pole in $f$ gives a logarithmic divergence.  Using the same belt geometry and assuming the surface drops to $z = 1 - \delta$, the area is
\begin{equation}
A_1 = 2 L_y \int_\epsilon^{1-\delta} \frac{dz}{z^2 \sqrt{f}} + L_y \frac{L_x}{(1-\delta)^2}.
\end{equation}
The first term diverges like $\ln{\delta}$ as $\delta \rightarrow 0$, and minimizing this formula with respect to $\delta$ gives $\delta \sim 1/L_x$ and a violation of the boundary law.

However, there is another smooth candidate, call it $\Sigma_2$, for the minimal surface.  The conserved quantity is now given by
\begin{equation}
\frac{1}{z^2} \frac{1}{\sqrt{1 + \dot{z}^2/f}} = \frac{1}{a^2}.
\end{equation}
The maximum value $a$ of $z$ still occurs at $x=0$ where $\dot{z} = 0$.  The area in terms of $a$ is given by
\begin{equation}
A_2 = 2 L_y \int_0^a dz \frac{1}{\sqrt{f}}\frac{a^2}{z^2 \sqrt{a^4 - z^4}},
\end{equation}
and the relationship between $a$ and $L_x$ is given by
\begin{equation}
L_x = 2\int_0^a dz \frac{1}{\sqrt{f}}\frac{z^2}{\sqrt{a^4 - z^4}}.
\end{equation}
These integrals can be solved numerically to find the relationship between $A_2$ and $L_x$.  Since the $a \rightarrow 1$ limit is singular, it is possible to verify that $A \sim L_y L_x$ as $L_x$ gets large.  This is an extensive contribution from the boundary point of view and corresponds to the large ground state degeneracy of the extremal black hole.  Numerical results confirm this extensive behavior and indicate that there is no sub-leading logarithmic correction in $L_x$ despite the similarity between $\Sigma_1$ and $\Sigma_2$ at large $L_x$.  Thus $A_2 < A_1$ due to the absence of the logarithm in $A_2$, and the minimal surface is $\Sigma_2$.  Following the holographic prescription, $\Sigma_2$ gives an entropy whose leading term as a function of $L_x$ is extensive with sub-leading corrections that obey the usual boundary law.

\begin{figure}
\includegraphics[width=\textwidth]{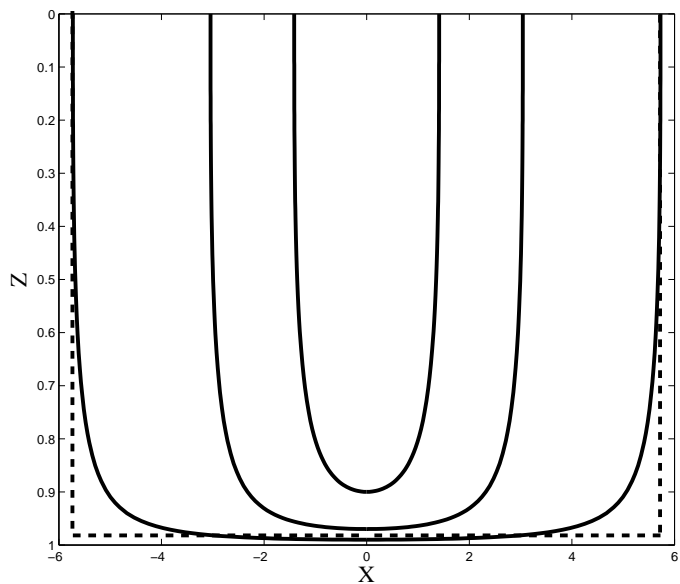}
\label{fig_1}
\caption{Cross sections at fixed $y$ of various minimal curves in the extremal case.  The horizon is at $z=1$.  The solid curves are three examples of a $\Sigma_2$ surface for various values of $L_x$.  The dashed curve falling directly to the near horizon is the corresponding $\Sigma_1$ curve for the largest value of $L_x$ shown.}
\end{figure}

\section{Conclusions}
I have computed the entanglement entropy of a $2+1$d quantum field theory at zero temperature and finite density via gauge/gravity duality.  According to the holographic dictionary the entanglement entropy is proportional to the area of a minimal surface in the bulk.  This prescription gives an extensive entropy at leading order due to the ground state degeneracy of the extremal black hole.  Subtracting this extensive piece, the contribution to the entropy from entanglement obeys the usual boundary law.  Despite some preliminary indications to the contrary, the true minimal surface does not give rise to a logarithmic correction to the boundary law in the dual field theory.

The original motivation was to demonstrate that a non-Fermi liquid with finite Fermi surface violates the boundary law for entanglement entropy.  However, I have found that such a violation is not encoded in the bulk geometry alone.  In order to verify that a violation does exist it would be nice to compute corrections to the entanglement entropy on the bulk side.  It has recently been possible to see quantum oscillations in the dyonic version of the extremal black hole by computing one loop corrections to the free energy \cite{bh_qo}.  A similar calculation of $1/N$ suppressed corrections should be possible for the entanglement entropy, perhaps by studying a fermion determinant in the Fursaev background.

\section{Acknowledgements}
I would like to thank John McGreevy, David Vegh, and Nabil Iqbal for helpful discussions.  This research was supported in part by Perimeter Institute for Theoretical Physics.

\bibliography{extremal_ee}

\end{document}